# Multiple and spectrally robust photonic magic angles in reconfigurable α-MoO$_3$ trilayers

J. Duan[1,2,†], G. Álvarez-Pérez[1,2], C. Lanza[1,2], A.I.F. Tresguerres-Mata[1,2], K. Voronin[3], N. Capote-Robayna[3], A. Tarazaga Martín-Luengo[1,2], J. Martín-Sánchez[1,2], V.S. Volkov[4,5], A.Y. Nikitin[3,6,†], and P. Alonso-González[1,2,†]

[1]Department of Physics, University of Oviedo, Oviedo 33006, Spain.

[2]Center of Research on Nanomaterials and Nanotechnology, CINN (CSIC-Universidad de Oviedo), El Entrego 33940, Spain.

[3]Donostia International Physics Center (DIPC), Donostia-San Sebastián 20018, Spain.

[4]Center for Photonics and 2D Materials, Moscow Institute of Physics and Technology, Dolgoprudny, 141700, Russia.

[5]GrapheneTek, Skolkovo Innovation Center, Moscow, Russia.

[6]IKERBASQUE, Basque Foundation for Science, Bilbao 48013, Spain.

†Corresponding to: duanjiahua@uniovi.es, alexey@dipc.org, pabloalonso@uniovi.es

**The assembling of twisted stacks of van der Waals (vdW) materials had led to the discovery of a profusion of remarkable physical phenomena in recent years, as it provides a means to accurately control and harness electronic band structures. This has given birth to the so-called field of twistronics. An analogous concept has been developed for highly confined polaritons, or nanolight, in twisted bilayers of strongly anisotropic vdW materials, extending the field to the twistoptics realm. In this case, the emergence of a topological transition of the polaritonic dispersion at a given twist angle (photonic magic angle) results in the propagation of nanolight along one specific direction (canalization regime), holding promises for unprecedented control of the flow of energy at the nanoscale. However, there is a fundamental limitation in twistoptics that critically impedes such control: there is only one photonic magic angle (and thus canalization direction) in a twisted bilayer and it is fixed for each incident frequency. Here, we overcome this limitation by demonstrating the existence of multiple spectrally robust photonic magic angles in reconfigurable twisted vdW trilayers. As a result, we show that canalization of nanolight can be programmed at will along any desired in-plane direction in a single device, and, importantly, within broad spectral ranges of up to 70 cm$^{-1}$. Our findings lay the**

**foundation for robust and widely tunable twistoptics, opening the door for applications in nanophotonics where on-demand control of energy at the nanoscale is crucial, such as thermal management, nanoimaging or entanglement of quantum emitters.**

**<u>Keywords:</u> phonon polaritons, twisted vdW layers, light canalization, photonic magic angle**

Twisted vdW materials exhibit a plethora of extraordinary physical phenomena, ranging from superconductivity[1,2], fractal quantum Hall effects[3-5] and ferromagnetism[6] in electronics, to moiré excitons[7-10], anomalous collective excitations[11], enhanced photoresponse[12,13], and nanoscale photonic crystals[14] in optics. In particular, the emerging field of twistoptics[15-20] seeks to steer on demand the propagation of nanolight – coupled photons to crystal lattice vibrations and referred to as phonon polaritons (PhPs)[21-23] – by exploiting twisted vdW heterostructures – thin in-plane anisotropic vdW crystal slabs rotated with respect to each other. Specifically, the observation of topological transitions in the dispersion of twisted bilayers of α-phase molybdenum trioxide (α-MoO$_3$)[16-19] has opened the door to an unprecedented control of their electromagnetic fields with inherent topological robustness[16]. The isofrequency curves (IFCs) - slices of the dispersion by a plane of constant frequency - of PhPs evolve from open (hyperbolic-like) to closed (elliptic-like) curves as a function of the twist angle. At the transition point, given by a "photonic magic angle" ($\theta_c$), the IFC flattens, giving rise to an exotic propagation state, the so-called canalization regime[24]. In this regime, nanolight propagates in a highly collimated manner, predominantly along a single spatial direction, with no geometrical decay and with an enhanced local density of states[25,26]. Polariton canalization[27], as well as the development of twistoptics, have thus great potential not only for a deeper understanding of fundamental topological and optical phenomena at the nanoscale, but also for applications such as waveguides, photonic circuits and infrared sensors. However, while a variety of extraordinary electronic phenomena have been studied in a large family of twisted vdW materials—such as magic-angle twisted bilayer graphene[1,28-30], twisted trilayer graphene[2] and twisted double bilayer graphene[31,32]—, up to date twisted α-MoO$_3$ bilayers remain the only known system in which a photonic magic angle and low-loss polariton canalization have been demonstrated. Importantly, twisted α-MoO$_3$ bilayers present a fundamental limitation: there exist only one photonic magic angle at each frequency ($\omega_c$). This limitation has important technological implications since for a determined twisted heterostructure both the propagation direction and wavelength of the canalized PhPs are intrinsically fixed by $\theta_c$ and the thickness of the α-MoO$_3$ layers. In addition, even when devices with other $\theta_c$ and thicknesses are available, the range of achievable canalization directions is much narrower than 0-180°. Indeed the propagation of PhPs is forbidden in a wide range of directions around the [001] crystal direction in α-MoO$_3$ in the hyperbolic *reststrahlen* band -spectral range in which the absorption through the crystal is strongly suppressed and the dielectric permittivity tensor $\hat{\varepsilon}$ has different signs of its elements-from 820 cm$^{-1}$ to 960 cm$^{-1}$ [33-35].

Here, we overcome this fundamental limitation of twistoptics by demonstrating the existence of multiple spectrally robust photonic magic angles in twisted α-MoO$_3$ trilayers, which enable on-demand canalization of PhPs along any spatial direction and within a wide spectral range that covers up to half (from 870 cm$^{-1}$ to 940 cm$^{-1}$) the hyperbolic *reststrahlen* band. To this end, we design and fabricate reconfigurable devices by stacking three single α-MoO$_3$ layers, precisely controlling the twist angles between them (see schematic in Fig. 1a,b and details in Methods). $\theta_{1-2}$ and $\theta_{1-3}$ are the twist angles between the [001] crystal direction of the bottom layer and the [001] crystal direction of the middle and top layers, respectively (all angles are defined as positive counterclockwise). Fig. 1c shows the optical image of a representative twisted α-MoO$_3$ trilayer device in which the top, middle and bottom layers are marked by white, yellow, and black dashed rectangles, respectively. In order to excite PhPs, we fabricate an Au rod antenna[36,37] between the top and middle layers. To visualize the propagation of PhPs, we perform near-field nanoimaging employing a scattering-type scanning near-field optical microscope (s-SNOM). As shown in Fig.1b, the tip of the s-SNOM and the sample are illuminated with s-polarized mid-infrared light spanning the hyperbolic *reststrahlen* band of α-MoO$_3$[35] under study.

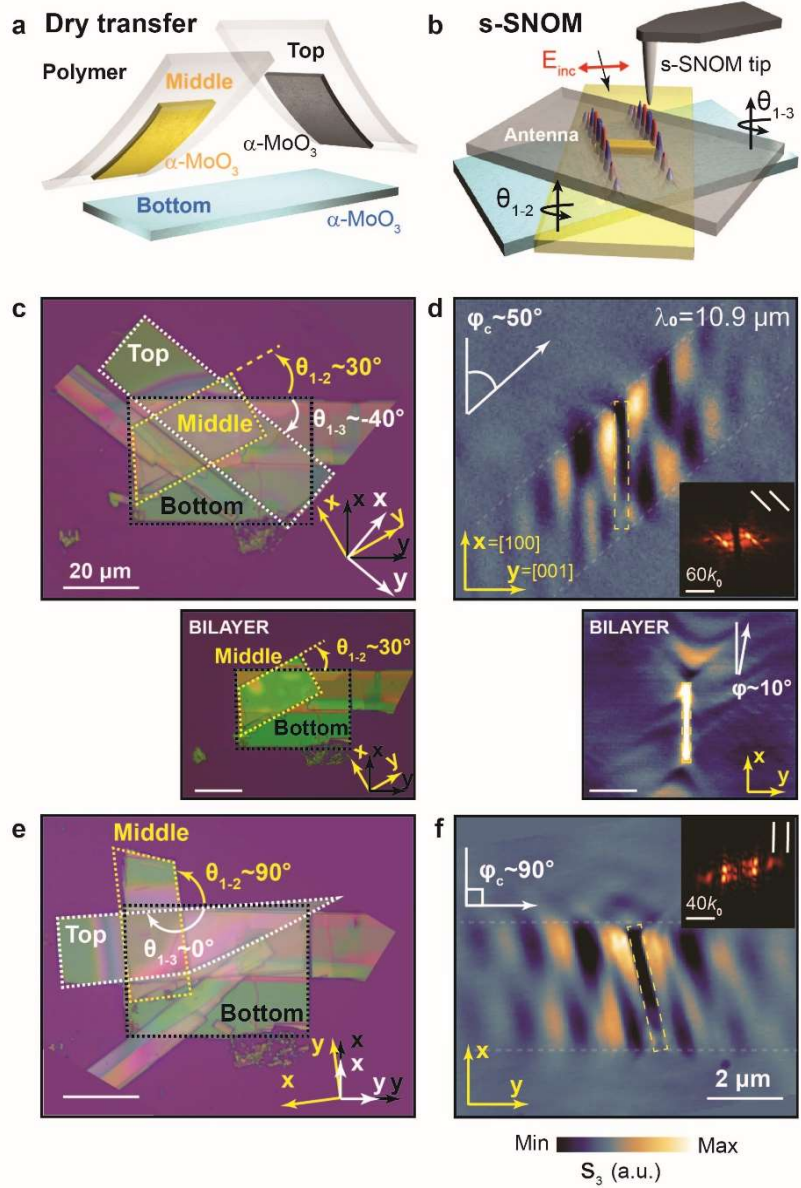

**Fig.1 Photonic magic angles and polariton canalization in reconfigurable trilayers. a,** Schematic of the dry-transfer fabrication of twisted trilayers, which can be reconfigured multiple times by disassembling and reassembling the layers at different twist angles. **b,** Schematic of the s-SNOM experimental scheme used to image PhPs in twisted α-MoO$_3$ trilayers. An Au antenna aligned along the [100] crystal direction of the middle α-MoO$_3$ layer launches the PhPs, whose wavefronts and propagation are probed with the s-SNOM tip. $\theta_{1\text{-}2}$ and $\theta_{1\text{-}3}$ are the twist angles between the [001] crystal direction of the bottom layer and the [001] crystal direction of the middle and top layers, respectively. The thickness of the top/middle/bottom layer is denoted by $d_1/d_2/d_3$. **c,** Top panel: optical image of a twisted α-MoO$_3$ trilayer ($d_1$ = 250nm, $d_2$ = 240nm, $d_3$ = 180nm) with $\theta_{1\text{-}2}$ = 30º, $\theta_{1\text{-}3}$ = -40º. The contour of the top, middle and bottom layer is marked by white, yellow, and black dashed rectangles, respectively. The $x$ and $y$ axes are defined along the [100] and [001] crystal directions of the α-MoO$_3$ layers. Bottom panel: optical image of the bilayer device ($\theta_{1\text{-}2}$ = 30º, $d_1$ = 240nm, $d_2$ = 180nm) intermediate between (c) and (e). **d,** Top panel: near-field amplitude image of the twisted trilayers in (c) at an illuminating wavelength of $\lambda_0$=10.9 μm. The PhPs launched by both extremities of the antenna are constrained within a narrow spatial sector along the direction $\varphi_c$=50º (white dashed lines). The experimental IFC (fast Fourier transform of the near-field image) is shown in the inset. The white solid lines are guides to the eye. Bottom panel: near-field amplitude image of the bilayer device in the bottom panel of (c) at an illuminating wavelength of $\lambda_0$=10.9 μm. **e**, Optical image of the same α-MoO$_3$ layers as in (c), but twisted at $\theta_{1-2}$ = 90º, $\theta_{1-3}$ = 0º. **f**, Near-field amplitude images of the twisted trilayers in (e) at $\lambda_0$=10.9 μm. The experimental IFC (inset) verifies polariton canalization along the direction of $\varphi_c$=90º.

As previously reported[16-19], α-MoO$_3$ bilayers exhibit canalization of PhPs when they are twisted at a single photonic magic angle. As such, at any other twist angle (see e.g. the bilayer with $\theta_{1-2}$ = 30º in Fig. 1c, bottom panel) the PhPs propagation is not canalized, ranging from isotropic to strongly anisotropic, such as the hyperbolic propagation shown in the near-field image of Fig. 1d (bottom panel) taken at an incident wavelength $\lambda_0$=10.9 μm. Excitingly, if we add a third layer of α-MoO$_3$ (Fig.1c) on top of this bilayer device (at $\theta_{1-3}$ = -40º), we observe that the propagation of PhPs changes dramatically, showing now a strongly canalized regime along the angle $\varphi_c$~50º (Fig.1d). This effect is better revealed by performing a fast Fourier transform (FFT) on the near-field image to extract the PhPs IFC (inset in Fig.1d), which exhibits two flat branches, unambiguously demonstrating the appearance of a canalization regime. Therefore, the addition of a third layer clearly extends the possibilities for canalization beyond those found in twisted bilayers. Interestingly, by disassembling the trilayer device (Fig.1c) into individual blocks and reassembling them into a new trilayer configuration with different twist angles, $\theta_{1-2}$ = 90º and $\theta_{1-3}$ = 0º (Fig.1e), we again observe PhPs canalization, which however, emerges along a different direction $\varphi_c$~90º (Fig.1f). Notably, such canalization direction corresponds to the [001] direction of the middle α-MoO$_3$ layer, which is forbidden for propagating PhPs in twisted bilayers at this frequency. These findings indicate that twisted trilayers allow the existence of multiple photonic magic angles and, importantly, the tuning of canalized PhPs, which opens the door to reconfigure and steer nanolight along any in-plane direction.

To explore the steering of polariton canalization, we calculate, both analytically and numerically, the IFCs of PhPs in α-MoO$_3$ trilayers twisted at a representative combination of twist angles. Note that finite thicknesses of α-MoO$_3$ layers have been considered in our theoretical models (see details in the Supplementary Information). Fig. 2a shows the resulting numerical IFCs for $\theta_{1-2}$ = 0º, 30º, 60º, 90º and $\theta_{1-3}$ = 0º, -40º, -60º, -90º (the variation of all angles from -90º to 90º and the analytical results can be found in the Supplementary Information), in excellent agreement with the analytical results (see

Fig. S5). Particularly, when $\theta_{1\text{-}3}$ varies from 0° to -90° for a fixed $\theta_{1\text{-}2}$ =0°, we observe a transition from open to closed IFCs (first column of Fig. 2a) passing through a flattened IFC that indicates the existence of a photonic magic angle. This "open-to-closed" topological transition is thus similar to the one reported in previous works for twisted bilayers of α-MoO$_3$[16-19]. On the other hand, when $\theta_{1\text{-}3}$ varies from 0° to -90° with a fixed $\theta_{1\text{-}2}$ =30° (second column of Fig. 2a) or a fixed $\theta_{1\text{-}2}$ =60° (third column of Fig.2 a), we observe that the IFCs are always flattened (except in the case of $\theta_{1\text{-}2}$ =30° and $\theta_{1\text{-}3}$ =0°, where the IFCs are open hyperbolas) but, interestingly, with slopes along different directions. This finding reveals the existence of multiple photonic magic angles at the same incident frequency in twisted trilayers. As a result, polaritons can be canalized along multiple directions $\varphi_c$ (white arrows in Fig. 2a) at the same frequency, depending upon the mutual orientation of the twisted layers in the heterostructure. For example $\varphi_c$~140° at $\theta_{1\text{-}2}$ =30° and $\theta_{1\text{-}3}$ =−90° (panel 1); $\varphi_c$~80° at $\theta_{1\text{-}2}$ =30° and $\theta_{1\text{-}3}$ =−60° (panel 2); or $\varphi_c$~50° at $\theta_{1\text{-}2}$ =30° and $\theta_{1\text{-}3}$ =−40° (panel 3). Interestingly, at the magic angles $\theta_{1\text{-}2}$ =90° and $\theta_{1\text{-}3}$ =0° (bottom panel of the fourth column of Fig. 2a) canalization of PhPs can be achieved along the direction $\varphi_c$~90°, and thus along the [001] main crystal axis. Note that for a single α-MoO$_3$ layer, no PhPs can propagate along this direction due to their hyperbolic dispersion (this particular case is shown experimentally in Fig. 1f).

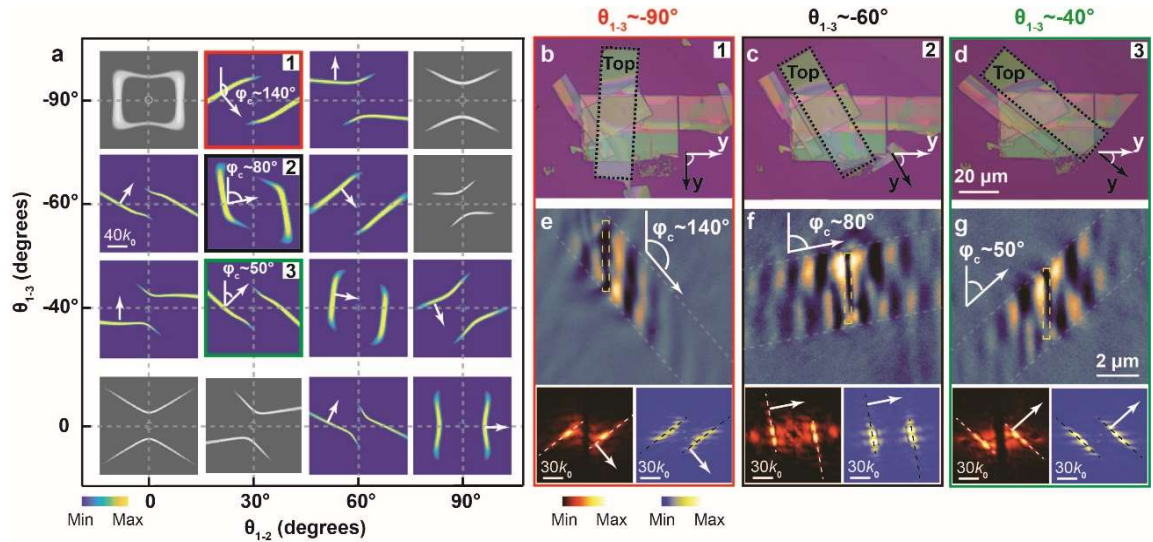

**Fig.2 Multiple photonic magic angles and tunable polariton canalization in twisted trilayers.**
**a,** Numerical IFCs for PhPs in a twisted α-MoO$_3$ trilayer with twist angles $\theta_{1–2}$ = 0°, 30°, 60°, 90° and $\theta_{1–3}$ = 0°, -40°, -60°, -90° (the variation of all angles from -90° to 90° can be found in Fig.S6). The panels with colored boxes (1,2,3) represent PhPs canalization regimes (i.e. flattened IFCs) along different directions (white arrows). Grey color plots represent non-canalized propagation regimes. **b-d,** Optical images of twisted trilayers for $\theta_{1–2}$ = 30° and $\theta_{1–3}$ = −90° (b), −60° (c), −40° (d), respectively. All samples are composed of the same three α-MoO$_3$ layers (d$_1$=250nm, d$_2$=240nm, d$_3$=180nm). **e-g,** Near-field amplitude images of the trilayer in (b), (c), and (d) at $\lambda_0$=10.9 μm. The propagation directions of PhPs are marked by white arrows, pointing along $\varphi_c$=140° (e), $\varphi_c$=80° (f), and $\varphi_c$=50° (g), respectively. The bottom left and right panels show the corresponding experimental (FFT of near-field images in (e-g)) and simulated (FFT of simulated images in Fig.S7) IFCs, respectively. The white and black dashed lines are guides to the eye. Images (e-g) correspond to the IFC shown in panels 1, 2, and 3 in (a), respectively. Canalization along other directions (all-angle tunability) is shown in Fig.S9.

Our theoretical calculations (Fig. 2a and Fig. S6) predict that PhPs in twisted α-MoO$_3$ trilayers can be designed to exhibit IFCs with much more diverse shapes (some of them even exhibiting strong asymmetries[38]) compared to twisted bilayers of the same vdW crystal, and in particular, with different flattening slopes, which can result in canalized PhPs along any direction in the plane (Fig. S9).

To experimentally demonstrate tuning of the PhP canalization, we performed near-field imaging of reconfigurable twisted α-MoO$_3$ trilayers in which we fixed the twist angle $\theta_{1-2}$ =30° and varied the angle $\theta_{1-3}$ multiple times (at -90° (Fig.2b), -60° (Fig. 2c) and -40° (Fig. 2d)). When $\theta_{1-3}$ = -90°, the resulting near-field image shows that the propagation of PhPs is strongly guided along a specific in-plane direction $\varphi_c$=140° (Fig. 2e). This is further verified by the experimental IFC (FFT of the near-field image), showing a flat contour formed by two parallel lines (left bottom panel in Fig.2e). In addition, we also performed numerical simulations mimicking the experiment (Fig. S7) and extracted the resulting IFC (FFT of the simulated electric field distribution), as shown in the right bottom panel of Fig. 2e. We find an excellent agreement between the experimental (left bottom panel in Fig. 2e) and simulated (right bottom panel in Fig. 2e) IFCs, which give rise to polariton canalization along the direction $\varphi_c$=140°. Interestingly, by changing the twist angle $\theta_{1-3}$ up to -60°, we observe that the canalization angle changes to $\varphi_c$=80° (Fig. 2f), yielding a flattened IFC (left bottom panel in Fig. 2f), in excellent agreement with the numerical simulations (right bottom panel in Fig. 2f). Varying $\theta_{1-3}$ further to -40°, we observe polariton canalization along another in-plane direction, $\varphi_c$=50° (Fig. 2g). These experimental results, together with our theoretical and numerical calculations, unequivocally demonstrate the existence of multiple photonic magic angles in twisted trilayers, which allow the propagation of canalized polaritons along any direction in the plane (see other possible canalization angles in Fig. S9). Note that the canalization in twisted trilayers give rise to low-loss (without geometrical decay, as shown in Fig.S10) and diffractionless propagation of nanolight, which is independent of the thickness of each layer (see details in Fig.S11). Furthermore, such steering capability in reconfigurable trilayers offers unprecedented possibilities for guiding mid-infrared electromagnetic fields at will at the nanoscale.

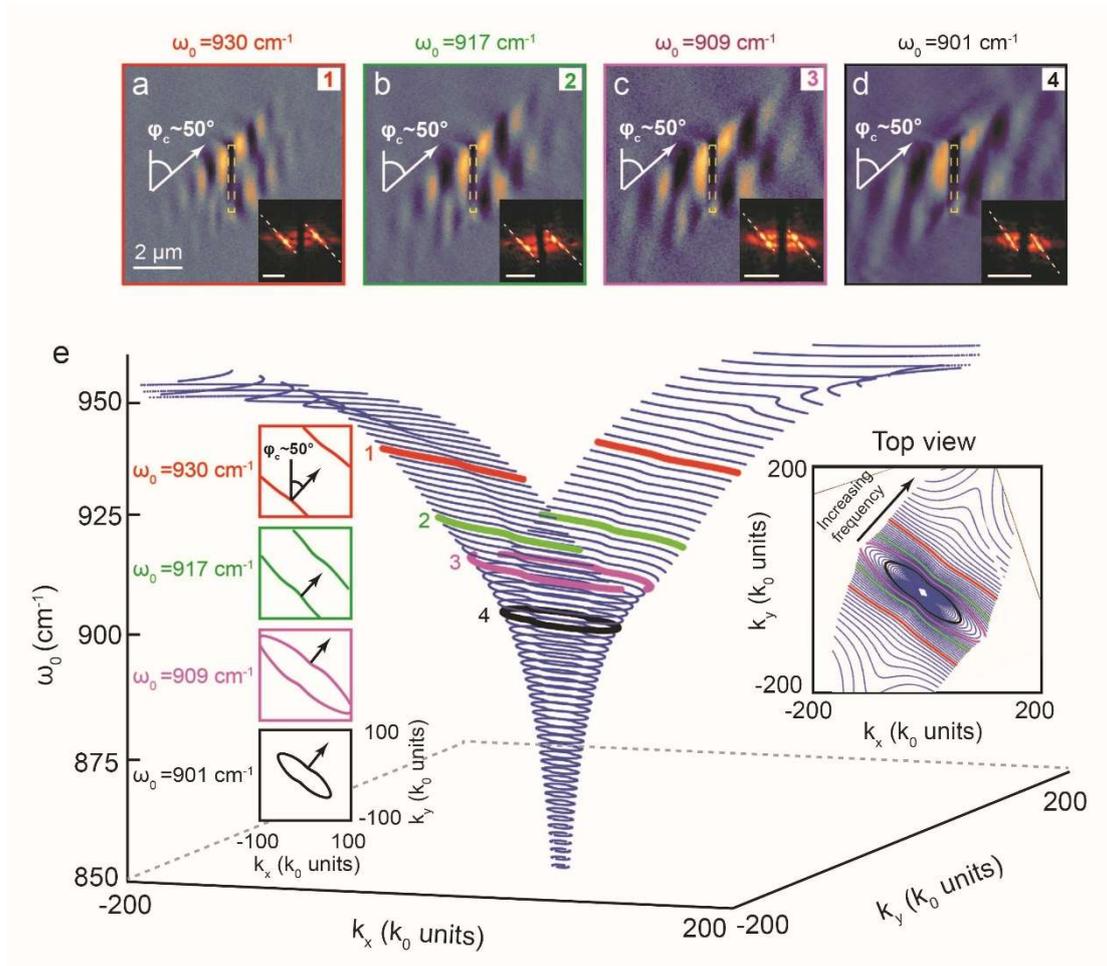

**Fig.3 Spectrally robust photonic magic angles and broadband polariton canalization in twisted trilayers. a-d,** Near-field amplitude images of a twisted trilayer device ($\theta_{1-2}$ = 30º, $\theta_{1-3}$ = -40º , $d_1$ = 250nm, $d_2$ = 240nm, $d_3$ = 180nm) at $\omega_0$=930 cm$^{-1}$ (b), $\omega_0$=917 cm$^{-1}$ (c), $\omega_0$=909 cm$^{-1}$ (d), $\omega_0$=901 cm$^{-1}$ (e), respectively. The propagation of PhPs is in all cases strongly collimated along a narrow spatial sector. The corresponding experimental IFCs (insets in (a-d)) features flat branches tilted along the same direction, verifying the emergence of one photonic magic angle within a wide spectral range, giving rise to canalization along the direction $\varphi_c$=50º (white arrows). The scale-bar in the FFT images is 50 $k_0$. **e,** Analytic IFCs of PhPs as a function of $\omega_0$ for the twisted trilayer shown in (a-d). Both the 3D and the top view (right inset) show flattened IFCs in a wide frequency range (870-940 cm$^{-1}$). The left insets show 2D cross sections of IFCs at the four incident frequencies used in a-d, showing polariton canalization along the direction $\varphi_c$=50º (black arrows).

However, to fully achieve such a goal there is still another obstacle that has not been addressed yet in twistoptics: the photonic magic angle emerges at a specific frequency ($\omega_c$) for each twisted bilayer device, thus lacking spectral robustness and thereby limiting their potential for technological applications requiring canalization at multiple frequencies (e.g. for broadband guiding of nanolight).

To study the spectral dependence of a photonic magic angle (and thus of canalization) in twisted trilayers, we performed near-field nanoimaging of a twisted α-MoO$_3$ trilayer at different incident frequencies ($\omega_0$) for a given combination of twist angles ($\theta_{1-2}$ =30º and $\theta_{1-3}$ =–40º). Excitingly, the obtained near-field images (Fig.3a-3d for $\omega_0$=930 cm$^{-1}$, $\omega_0$=917 cm$^{-1}$, $\omega_0$=909 cm$^{-1}$, and $\omega_0$=901 cm$^{-1}$, respectively) show that the propagation of PhPs in the twisted trilayer is strongly collimated along a specific direction ($\varphi_c$=50º) in all cases, i.e., for a wide range of $\omega_0$. This result is further verified by transforming these images into IFCs, which show strongly flattened regions in all cases (insets of Fig.3b-3e), thus unambiguously demonstrating the robustness of a photonic magic angle in a twisted trilayer, giving rise to canalized polaritons along a specific direction in a broadband frequency range. To corroborate our experimental findings theoretically, we calculate the analytic IFCs as a function of the incident frequency ($\omega_0$) for the twisted trilayer shown in Fig. 3a-3d (Fig. 3e). In both the 3D view and the contour plot (inset of Fig.3e), the analytic IFCs show a topological transition from an elongated ellipse to an open curve with a rather complex contour. Importantly, within this transition the IFCs remain flattened along the same direction within a wide frequency range (870-940 cm$^{-1}$) covering up to half of the hyperbolic *reststrahlen* band of α-MoO$_3$ under study, thus revealing the emergence of a photonic magic angle with spectral robustness in twisted trilayers. This is better observed by two-dimensional (2D) cross sections at the four representative frequencies shown in Fig. 3a-3d (red, green, purple, and black squares in the inset of Fig.3e for $\omega_0$=930 cm$^{-1}$, $\omega_0$=917 cm$^{-1}$, $\omega_0$=909 cm$^{-1}$, and $\omega_0$=901 cm$^{-1}$, respectively), which show in all cases IFCs flattened along the same direction $\varphi_c$=50º (black arrows). Our analytic calculations are thus in excellent agreement with our experimental results, corroborating the existence of robust photonic magic angles in twisted α-MoO$_3$ trilayers.

Overall, our findings demonstrate the existence of multiple photonic magic angles in reconfigurable twisted α-MoO$_3$ trilayers, which enable on-demand polariton canalization along an arbitrary in-plane direction by simply varying the interlayer twist angles. Furthermore, we find that these photonic magic angles are spectrally robust, i.e., they arise within a wide frequency range, giving rise to broadband canalization of nanolight, crucial for the implementation of twisted vdW heterostructures in optical nanotechnologies. Interestingly, we envision that even wider spectral ranges could potentially be covered by incorporating additional twisted layers of α-MoO$_3$ or other vdW materials supporting polaritons, such as V$_2$O$_5$ and intercalated-V$_2$O$_5$[39], WTe$_2$[40], or graphene[41-43]. Altogether, our results bring twistoptics to the next level, opening a plethora of new possibilities for controlling and steering the flow of electromagnetic energy at will at the nanoscale.

## Methods

**Fabrication of reconfigurable twisted trilayer devices.**

Reconfigurable twisted α-MoO$_3$ trilayer devices were fabricated using the dry transfer technique[44]. First, we performed a mechanical exfoliation from commercial α-MoO$_3$ bulk materials (Alfa Aesar) using Nitto tape (Nitto Denko Co., SPV 224P). Next, homogeneous α-MoO$_3$ flakes were exfoliated from the tape to a transparent poly-(dimethylsiloxane) (PDMS) stamp and selected with the desired thickness by an optical microscope. Then, we applied a home-made micromanipulator for aligning and twisting the PDMS stamps with α-MoO$_3$ flakes. Once the first α-MoO$_3$ flake was released on a SiO$_2$ substrate by heating the PDMS stamp to 200 °C, the second and third α-MoO$_3$ flake were precisely aligned, twisted at the desired angle, and released one by one on top of the first flake. After that, we disassembled the as-fabricated trilayer devices into individual blocks. To do that, we picked up the top α-MoO$_3$ flake using a stamp formed by a PDMS film covered with a layer of polypropylene carbonate (PC)[45] and repeated the same process for the middle α-MoO$_3$ flake. Finally, we precisely aligned the PC/PDMS stamp with desired twist angles, reassembled the middle and bottom α-MoO$_3$ layers, heated them up to 200 °C to peel off the PDMS from the PC, and removed the PC layer by immersing the sample into chloroform solvent at 100 °C. The same process was repeated with the top α-MoO$_3$ layer. It is noted that the disassembling and reassembling process can be easily performed multiple times, resulting in a reconfigurable twisted trilayer device in which the twist angles can be changed multiple times.

**Fabrication of gold antennas.**

We fabricated gold antennas with the dimensions of 3 μm (length) × 50 nm (width) × 40 nm (height) on top of the middle α-MoO$_3$ layers. The longitudinal direction of the gold antenna was aligned along the [100] crystal direction of the middle α-MoO$_3$ layer. We performed a high-resolution electron beam lithography (100 kV and 100 pA) on the samples coated with a poly- (methyl methacrylate) (PMMA) resist layer. With conventional high-resolution developer (1:3 methyl isobutyl ketone (MIBK)/isopropyl alcohol (IPA)), evaporation of 5 nm Cr and 35 nm Au and a lift-off were done to define the antennas. After that, we immersed the sample into acetone at 60 °C for 10-15 mins and gently rinsed it with isopropyl alcohol (IPA) to remove organic residual, followed by a nitrogen gas drying and thermal evaporation.

**Near-field nanoimaging.**

All nanoimaging measurements (Fig1-3) were performed using a commercial scattering-type scanning near-field optical microscope (s-SNOM) from Neaspec GmbH. The s-SNOM is equipped with a quantum cascade laser (Daylight Solutions) spanning from 890-1140 cm$^{-1}$ and is based on an atomic force microscope (AFM) operating in the tapping mode with an amplitude of ~120 nm at $\Omega \approx 280$ kHz. Commercial metal-coated (Pt-Ir) AFM tips (ARROW-NCPt-50, Nanoworld) were employed. The gold antenna (length: 3μm, width: 40nm, height: 50nm) was illuminated with s-polarized mid-infrared light, with the electric field parallel to the longitudinal direction of the antenna. The tip-scattered light was collected by a parabolic mirror toward an infrared detector (Kolmar Technologies) enabling the direct visualization of the PhP wavefronts. The detected signal was demodulated at the third harmonic to suppress the far-field background scattering, while both the amplitude and phase signal were extracted based on a pseudoheterodyne interferometry.

**Numerical simulations.**

We used the finite boundary elements method in full-wave simulations based on commercially available software (COMSOL Multiphysics). We calculate the vertical component of the electric field ($|E_z|$) at 5 nm on top of the uppermost surface of the twisted α-MoO$_3$ trilayers (Fig.S7), which is then fast Fourier transformed to extract the simulated PhPs IFCs in the right bottom panels of Fig.2e-2g. The twisted α-MoO$_3$ trilayers were modeled as three biaxial layers rotated with respect to each other by the angles $\theta_{1-2}$ and $\theta_{1-3}$ on top of a SiO$_2$ substrate. PhPs were launched by gold rod antennas, whose geometric parameters are the same as the ones used in the s-SNOM measurements. The gold antennas were placed between the top and middle α-MoO$_3$ layers and illuminated by far-field radiation polarized along their longitudinal axis. The thicknesses of α-MoO$_3$ layers had the same value as in the experiments: d1=250nm, d2=240nm, d3=180nm. The dielectric permittivities for α-MoO$_3$ and SiO$_2$ have been taken from Ref[35] and Ref[46], respectively. The numerical IFCs were obtained by transfer matrix methods[47], where the imaginary part of the Fresnel reflection coefficient of the twisted trilayer systems was calculated (See details in the supplementary information).

**Analytical calculations.**

We developed three increasingly sophisticated analytical models for calculating the PhP polaritons in the twisted trilayers of biaxial crystals: (i) 2D-approximation model (Part A in the Section S1 of the Supplementary Information); (ii) large-$k$-approximation model (Part B in the Section S1 of the Supplementary Information); (iii) large-$k$-approximation model with air gaps (Part C in the Section S1 of the Supplementary Information). In the large-$k$-approximation model (ii), we considered the top/middle/bottom layers of a finite thickness and derived the implicit dispersion relation:

$$\tan(q_{1z}k_0 d_1)\tan(q_{2z}k_0 d_2)\tan(q_{3z}k_0 d_3)[\varepsilon_1\varepsilon_2 q^2 q_{2z}^2 - \varepsilon_z^2 q_{1z}^2 q_{3z}^2]$$
$$+ \tan(q_{1z}k_0 d_1)\tan(q_{2z}k_0 d_2)\varepsilon_z q q_{3z}(\varepsilon_1 q_{2z}^2 + \varepsilon_2 q_{1z}^2)$$
$$+ \tan(q_{1z}k_0 d_1)\tan(q_{3z}k_0 d_3)\varepsilon_z q q_{2z}(\varepsilon_1 q_{3z}^2 + \varepsilon_2 q_{1z}^2)$$
$$+ \tan(q_{2z}k_0 d_2)\tan(q_{3z}k_0 d_3)\varepsilon_z q q_{1z}(\varepsilon_1 q_{3z}^2 + \varepsilon_2 q_{2z}^2)$$
$$+ \tan(q_{1z}k_0 d_1)q_{2z}q_{3z}[\varepsilon_z^2 q_{1z}^2 - \varepsilon_1\varepsilon_2 q^2]$$
$$+ \tan(q_{2z}k_0 d_2)q_{1z}q_{3z}[\varepsilon_z^2 q_{2z}^2 - \varepsilon_1\varepsilon_2 q^2]$$
$$+ \tan(q_{1z}k_0 d_3)q_{1z}q_{2z}[\varepsilon_z^2 q_{3z}^2 - \varepsilon_1\varepsilon_2 q^2] - (\varepsilon_1+\varepsilon_2)\varepsilon_z q q_{1z}q_{2z}q_{3z} = 0,$$

Where $\varepsilon_1$ and $\varepsilon_2$ are the dielectric permittivities of the superstrate and substrate, respectively. $\varepsilon_z$ is the out-of-plane dielectric permittivity of the biaxial slabs. d1, d2, and d3 represent the thickness of top, middle, and bottom layer, respectively. $q_{iz}$ ($i$=1, 2, 3) stands for the out-of-plane component of the wavevector of the $i$-th layer:

$$q_{iz} = \sqrt{-\frac{\varepsilon_{ix}}{\varepsilon_{iz}}q_{ix}^2 - \frac{\varepsilon_{iy}}{\varepsilon_{iz}}q_{iy}^2},$$

where $\varepsilon_{ix}$, $\varepsilon_{iy}$ represent the in-plane components of the dielectric permittivity of the $i$-th ($i$=1, 2, 3) biaxial slabs, while $\varepsilon_{iz}$ represent its out-of-plane component. $q_{ix}$ and $q_{iy}$ are the in-plane component of the wavevector of the $i$-th slab. This model does not take into account air gaps between the slabs, leading to a slight difference between the analytical IFCs (Fig.S3) and experimental results (Fig.2). A more sophisticated model (iii) (large-$k$-air-approximation model) with the inclusion of air gaps (see details in the Section 1 of the Supplementary Information), allowed us to achieve a perfect agreement with our experimental results (Fig.S5).

**Author contributions**

J.D. and P.A.-G. conceived the study. P.A.-G. and A.Y.N. supervised the project. J.D. and A.I.F.T-M. fabricated the twisted samples. J.D. carried out the near-field imaging experiments. C.L., K.V., and G.A.-P. carried out numerical simulations and analytical calculations with the help of N.C.-R. and A.T.M.-L. J.D., P.A.-G., A.Y.N., J.M.-S., V.S.V. and G.A.-P. participated in data analysis. J.D., P.A.-G., G.A.-P. and C.L. co-wrote the manuscript with input from the rest of authors.

**Acknowledgements**

A.I.F.T.-M. and G.A-P. acknowledge support through the Severo Ochoa program from the Government of the Principality of Asturias (nos. PA-21-PF-BP20-117 PA-20-PF-BP19-053, respectively). J.M.-S. acknowledges financial support from the Ramón y Cajal Program of the Government of Spain and FSE (RYC2018-026196-I) and the Spanish Ministry of Science and Innovation (State Plan for Scientific and Technical Research and Innovation grant number PID2019-110308GA-I00). P.A.-G. acknowledges support from the European Research Council under starting grant no. 715496, 2DNANOPTICA and the Spanish Ministry of Science and Innovation (State Plan for Scientific and Technical Research and Innovation grant number PID2019-111156GB-I00). A.Y.N. acknowledges the Spanish Ministry of Science and Innovation (grant PID2020-115221GB-C42) and the Basque Department of Education (grant PIBA-2020-1-0014).

# References

1. Cao, Y. *et al.* Unconventional superconductivity in magic-angle graphene superlattices. *Nature* **556**, 43-50 (2018).
2. Park, J. M., Cao, Y., Watanabe, K., Taniguchi, T. & Jarillo-Herrero, P. Tunable strongly coupled superconductivity in magic-angle twisted trilayer graphene. *Nature* **590**, 249-255 (2021).
3. Hunt, B. *et al.* Massive Dirac fermions and Hofstadter butterfly in a van der Waals heterostructure. *Science* **340**, 1427-1430 (2013).
4. Dean, C. R. *et al.* Hofstadter's butterfly and the fractal quantum Hall effect in moiré superlattices. *Nature* **497**, 598-602 (2013).
5. Ponomarenko, L. *et al.* Cloning of Dirac fermions in graphene superlattices. *Nature* **497**, 594-597 (2013).
6. Sharpe, A. L. *et al.* Emergent ferromagnetism near three-quarters filling in twisted bilayer graphene. *Science* **365**, 605-608 (2019).
7. Jin, C. *et al.* Observation of moiré excitons in WSe2/WS2 heterostructure superlattices. *Nature* **567**, 76-80 (2019).
8. Seyler, K. L. *et al.* Signatures of moiré-trapped valley excitons in MoSe2/WSe2 heterobilayers. *Nature* **567**, 66-70 (2019).
9. Tran, K. *et al.* Evidence for moiré excitons in van der Waals heterostructures. *Nature* **567**, 71-75 (2019).
10. Alexeev, E. M. *et al.* Resonantly hybridized excitons in moiré superlattices in van der Waals heterostructures. *Nature* **567**, 81-86 (2019).
11. Hesp, N. C. *et al.* Observation of interband collective excitations in twisted bilayer graphene. *Nature Physics* **17**, 1162-1168 (2021).
12. Sunku, S. S. *et al.* Hyperbolic enhancement of photocurrent patterns in minimally twisted bilayer graphene. *Nature communications* **12**, 1-7 (2021).
13. Hesp, N. C. *et al.* Nano-imaging photoresponse in a moiré unit cell of minimally twisted bilayer graphene. *Nature communications* **12**, 1-8 (2021).
14. Sunku, S. *et al.* Photonic crystals for nano-light in moiré graphene superlattices. *Science* **362**, 1153-1156 (2018).
15. Hu, G., Krasnok, A., Mazor, Y., Qiu, C.-W. & Alù, A. Moiré hyperbolic metasurfaces. *Nano Letters* **20**, 3217-3224 (2020).
16. Hu, G. *et al.* Topological polaritons and photonic magic angles in twisted α-MoO3 bilayers. *Nature* **582**, 209-213 (2020).
17. Chen, M. *et al.* Configurable phonon polaritons in twisted α-MoO3. *Nature Materials* **19**, 1307-1311 (2020).
18. Duan, J. *et al.* Twisted nano-optics: manipulating light at the nanoscale with twisted phonon polaritonic slabs. *Nano Letters* **20**, 5323-5329 (2020).
19. Zheng, Z. *et al.* Phonon polaritons in twisted double-layers of hyperbolic van der Waals crystals. *Nano letters* **20**, 5301-5308 (2020).
20. Herzig Sheinfux, H. & Koppens, F. H. The rise of twist-optics. *Nano Letters* **20**, 6935-6936 (2020).
21. Basov, D., Fogler, M. & García de Abajo, F. Polaritons in van der Waals materials. *Science* **354**, aag1992 (2016).
22. Low, T. *et al.* Polaritons in layered two-dimensional materials. *Nature materials* **16**, 182-194 (2017).
23. Caldwell, J. D. *et al.* Low-loss, infrared and terahertz nanophotonics using surface phonon polaritons. *Nanophotonics* **4**, 44-68 (2015).

24	Li, P. *et al.* Collective near-field coupling and nonlocal phenomena in infrared-phononic metasurfaces for nano-light canalization. *Nature communications* **11**, 1-8 (2020).
25	Gomez-Diaz, J. S., Tymchenko, M. & Alu, A. Hyperbolic plasmons and topological transitions over uniaxial metasurfaces. *Physical review letters* **114**, 233901 (2015).
26	Krishnamoorthy, H. N., Jacob, Z., Narimanov, E., Kretzschmar, I. & Menon, V. M. Topological transitions in metamaterials. *Science* **336**, 205-209 (2012).
27	Zhang, Q. *et al.* Interface nano-optics with van der Waals polaritons. *Nature* **597**, 187-195 (2021).
28	Tomarken, S. L. *et al.* Electronic compressibility of magic-angle graphene superlattices. *Physical review letters* **123**, 046601 (2019).
29	Andrei, E. Y. & MacDonald, A. H. Graphene bilayers with a twist. *Nature materials* **19**, 1265-1275 (2020).
30	Yankowitz, M. *et al.* Tuning superconductivity in twisted bilayer graphene. *Science* **363**, 1059-1064 (2019).
31	Rubio-Verdú, C. *et al.* Moiré nematic phase in twisted double bilayer graphene. *Nature Physics* **18**, 196-202 (2022).
32	Shen, C. *et al.* Correlated states in twisted double bilayer graphene. *Nature Physics* **16**, 520-525 (2020).
33	Ma, W. *et al.* In-plane anisotropic and ultra-low-loss polaritons in a natural van der Waals crystal. *Nature* **562**, 557-562 (2018).
34	Zheng, Z. *et al.* A mid-infrared biaxial hyperbolic van der Waals crystal. *Science advances* **5**, eaav8690 (2019).
35	Álvarez‐Pérez, G. *et al.* Infrared permittivity of the biaxial van der waals semiconductor $\alpha$-MoO3 from near‐and far‐field correlative studies. *Advanced Materials* **32**, 1908176 (2020).
36	Alonso-González, P. *et al.* Controlling graphene plasmons with resonant metal antennas and spatial conductivity patterns. *Science* **344**, 1369-1373 (2014).
37	Pons-Valencia, P. *et al.* Launching of hyperbolic phonon-polaritons in h-BN slabs by resonant metal plasmonic antennas. *Nature communications* **10**, 1-8 (2019).
38	Passler, N. C. *et al.* Hyperbolic shear polaritons in low-symmetry crystals. *Nature* **602**, 595-600 (2022).
39	Taboada-Gutiérrez, J. *et al.* Broad spectral tuning of ultra-low-loss polaritons in a van der Waals crystal by intercalation. *Nature materials* **19**, 964-968 (2020).
40	Wang, C. *et al.* Van der Waals thin films of WTe2 for natural hyperbolic plasmonic surfaces. *Nature communications* **11**, 1-9 (2020).
41	Álvarez-Pérez, G. *et al.* Active tuning of highly anisotropic phonon polaritons in van der Waals crystal slabs by gated graphene. *ACS Photonics* **9**, 383-390 (2022).
42	Zeng, Y. *et al.* Tailoring topological transition of anisotropic polaritons by interface engineering in biaxial crystals. *arXiv preprint arXiv:2201.01412* (2022).
43	Hu, H. *et al.* Doping-driven topological polaritons in graphene/{\alpha}-MoO3 heterostructures. *arXiv preprint arXiv:2201.00930* (2022).
44	Castellanos-Gomez, A. *et al.* Deterministic transfer of two-dimensional materials by all-dry viscoelastic stamping. *2D Materials* **1**, 011002 (2014).
45	Frisenda, R. *et al.* Recent progress in the assembly of nanodevices and van der Waals heterostructures by deterministic placement of 2D materials. *Chemical Society Reviews* **47**, 53-68 (2018).

46  Aguilar-Merino, P. *et al.* Extracting the infrared permittivity of SiO2 substrates locally by near-field imaging of phonon polaritons in a van der Waals crystal. *Nanomaterials* **11**, 120 (2021).
47  Passler, N. C. & Paarmann, A. Generalized 4× 4 matrix formalism for light propagation in anisotropic stratified media: study of surface phonon polaritons in polar dielectric heterostructures. *JOSA B* **34**, 2128-2139 (2017).